\documentclass[sigconf]{acmart}

\usepackage{algorithm}
\usepackage{algorithmic}
\usepackage{comment}
\usepackage{adjustbox}
\usepackage{bm}
\usepackage{multirow}
\usepackage{siunitx}
\usepackage{placeins}
\usepackage{float}
\AtBeginDocument{%
  }

\setcopyright{acmlicensed}
\copyrightyear{2027}
\acmYear{2027}
\acmDOI{XXXXXXX.XXXXXXX}
\acmConference[Conference acronym 'XX]{Make sure to enter the correct
  conference title from your rights confirmation email}{June 03--05,
  2027}{Woodstock, NY}
\acmISBN{978-1-4503-XXXX-X/2027/06}

\begin{document}

\title{Neural Parameter Estimation of RC Thermal Building Models for Model Predictive Control}

\author{Fabian Raisch}
\orcid{1234-5678-9012}
\affiliation{%
 \institution{Technical University of Applied Sciences Rosenheim}
  \city{Rosenheim}
  \country{Germany}}
\affiliation{%
  \institution{Technical University of Munich}
  \city{Munich}
  \country{Germany}}

\author{Timo Germann}
\affiliation{%
 \institution{Technical University of Applied Sciences Rosenheim}
  \city{Rosenheim}
  \country{Germany}}
\affiliation{%
  \institution{Ludwig Maximilian University of Munich}
  \city{Munich}
  \country{Germany}}

\author{Sang-Woo Ham}
\affiliation{%
  \institution{Lawrence Berkeley National Laboratory}
  \city{Berkeley}
  \country{USA}}

\author{J. Nathan Kutz}
\affiliation{%
  \institution{Autodesk Research}
  \city{London}
  \country{UK}}

\author{Christoph Goebel}
\affiliation{%
  \institution{Technical University of Munich}
  \city{Munich}
  \country{Germany}}

\author{Benjamin Tischler}
\affiliation{%
 \institution{Technical University of Applied Sciences Rosenheim}
  \city{Rosenheim}
  \country{Germany}}

\renewcommand{\shortauthors}{Raisch et al.}

\sloppy
\begin{abstract}

Gray-box RC models are widely used to enable energy-efficient model predictive control (MPC) in buildings. However, estimating RC parameters remains difficult, as conventional optimization-based algorithms are prone to local minima, rely heavily on good initial guesses, and incur high computational cost. To address these issues, we propose the Estimator from Scratch, a novel neural parameter estimation approach that embeds the physical equations into a neural network's training process to estimate RC parameters. To further improve estimation accuracy and eliminate dependence on an initial guess, we extend this approach by pretraining the neural network on data from multiple source buildings, the Pretrained Estimator. We benchmark both methods against a genetic-algorithm-based RC estimator and a fully black-box neural network. All methods are evaluated across eight simulated and three real-world buildings for two RC configurations, for both prediction accuracy and closed-loop MPC performance, the latter only for the simulated buildings. The Pretrained Estimator achieves the best prediction performance among all RC-based methods, particularly with little training data, and yields the lowest and least variable MPC costs across buildings and benchmarks. These results position the Pretrained Estimator as a robust, computationally efficient, and initial-guess-free alternative for RC parameter estimation, with potential to extend to other control-oriented dynamical systems.

\end{abstract}




\maketitle

\section{Introduction}

Buildings are responsible for approximately one third of global CO$_2$ emissions and nearly 40\% of final energy consumption worldwide \cite{iea2022buildings}. A promising pathway to reduce these emissions is to equip buildings with advanced control based on model predictive control (MPC), which can reduce operational emissions by $10$--$50\%$ \cite{Drgona.2020}. However, MPC depends on an accurate model of the building's thermal dynamics. 
Approaches to modeling building dynamics range from white-box models based on detailed physical principles to purely black-box, data-driven models. White-box models are interpretable but require expensive, labor-intensive manual modeling. Black-box approaches are more scalable but offer limited interpretability, no physical guarantees, and typically require large, well-excited datasets for training \cite{Drgona.2020, mpc_excite}. Gray-box modeling combines both, with the Resistance--Capacitance (RC) network as its most common representation of building thermal dynamics \cite{SERASINGHE2024114123}. RC models are interpretable, data-efficient, and well-suited for control-oriented tasks such as MPC \cite{arroyo_identification_2020, SERASINGHE2024114123}. However, estimating the resistances and capacitances from measured data is a non-convex inverse problem \cite{drgona_physics-constrained_2021, arendt_modestpy_2019}, and conventional optimization-based estimation therefore faces recurring issues: (1) numerous local minima, (2) heavy dependence on a good initial guess, (3) sensitivity to the diversity of dynamics captured within the training data, and (4) large computational optimization cost \cite{arendt_modestpy_2019, arroyo_identification_2020, BLUM2019410, SERASINGHE2024114123}. These factors jointly limit the reliability, accuracy, and scalability of RC-based gray-box modeling. 

To address these limitations, we propose a neural parameter estimation approach for RC models, inspired by recent advances in applied mathematics \cite{doi:10.1073/pnas.2216415120}. This new paradigm maps time series data directly to parameter estimates via a neural network: the parameter space is explored using randomly sampled data windows while simultaneously adjusting the neural network weights, narrowing down the plausible parameter space over successive training windows. Gaskin et al.~\cite{doi:10.1073/pnas.2216415120} showed this to be faster and more accurate than classical methods such as Markov Chain Monte Carlo, making it a promising candidate for the non-convex RC estimation problem. However, their approach is limited to dynamical systems without control signals and disturbances.
To this end, we extend this approach to control-oriented settings and apply it to RC building models. We call the resulting method the \textit{Estimator from Scratch}. This gives rise to our first research question: \textit{Can the Estimator from Scratch achieve more accurate or similar parameter estimates than conventional optimization-based estimation for building RC models, while addressing some of their core issues?}

Neural parameter estimation, as suggested in \cite{doi:10.1073/pnas.2216415120}, relies on an initialization scheme in which the neural network is pretrained on randomly sampled inputs to statically predict an initial guess. Its performance can therefore be affected by the quality of this initial guess, similar to conventional optimization-based estimation \cite{SERASINGHE2024114123}. The transfer learning (TL) domain addresses this initialization problem by reusing a model pretrained on a source task for a target task. As a result, initialization procedures are omitted, and data requirements in the target are usually reduced \cite{pinto2022transfer, peirelinck2022transfer}. We therefore adopt this TL procedure and pretrain the neural network used for parameter estimation on source-building data. This yields the \textit{Pretrained Estimator}: a neural network that encodes generalizable knowledge from source buildings in its weights before it ever sees the target building data, thereby omitting the need for an initial guess.
This leads to our second research question: \textit{Can the Pretrained Estimator improve upon the Estimator from Scratch while removing the dependence on an initial guess?} This question is especially interesting for real-world buildings, since the Pretrained Estimator is pretrained on simulated data, raising a SIM2REAL problem.

For evaluation, we benchmark the Estimator from Scratch and the Pretrained Estimator against a Genetic Algorithm (GA), the most commonly used conventional RC estimation method \cite{SERASINGHE2024114123}, across two RC configurations. We further compare against a black-box neural baseline, a multilayer perceptron (MLP): first, to test whether the neural gray-box structure justifies its added complexity over its fully black-box alternative; second, because it has been found to perform best for modeling and control among several modeling paradigms \cite{stoffel2023evaluation}. All methods are evaluated via prediction accuracy across multiple training data lengths on eight simulated and three real-world buildings. As good prediction accuracy does not necessarily translate into good control performance \cite{BLUM2019410, martin_poster, mpc_excite}, we further evaluate all modeling approaches via closed-loop MPC on the simulated buildings and additionally compare against a PI controller.

The main contributions of this paper are:
\begin{itemize}
    \item Extending neural parameter estimation \cite{doi:10.1073/pnas.2216415120} to control-oriented dynamical systems and applying it to RC building models. 
    \item Neural parameter estimation achieves the best prediction accuracy in 20 of 22 building-RC pairs, while requiring lower estimation cost and showing less sensitivity to training length than the GA baseline.
    \item The Pretrained Estimator achieves the lowest MPC cost with the least variation among target buildings, outperforming MPC with a GA-estimated model by 12.9\% and a PI controller by 33.4\% on average.
\end{itemize}
The remainder of this paper is structured as follows: Section~\ref{ch:literature} reviews related work. Section~\ref{ch:method} describes our method and Section~\ref{sec_setup} the evaluation setup. Section~\ref{ch:experiments} presents the experiments, followed by the discussion in Section~\ref{ch:discussion} and the conclusion in Section~\ref{ch:conclusion}.

\section{Related Work}
\label{ch:literature}

Thermal dynamics models are a prerequisite for energy-efficient MPC in buildings \cite{Drgona.2020}. Therefore, Stoffel et al.~\cite{stoffel2023evaluation} compare white-box, RC-gray-box, and black-box models for MPC on a simulated one-zone office building. They find that a black-box MLP-based controller achieves the highest energy savings (8.4\%) compared to a rule-based controller, closely followed by the white-box (7.4\%) and RC-gray-box (7.2\%) variants. In a follow-up study, Stoffel et al. \cite{stoffel_real-life_2024} focus exclusively on black-box models, comparing a linear model, an MLP, and Gaussian Process regression in a real building. Here, the linear model achieves the best control performance, while the MLP achieves the best prediction performance. However, the linear model requires daily online updates, which are computationally expensive, and, as a black-box model, offers only limited physical interpretability. 

RC models, in contrast, can likewise be formulated in a linear fashion while remaining physically interpretable. Furthermore, RC models are data efficient and simple to integrate into MPC workflows, making them widely used in the literature \cite{SERASINGHE2024114123}. 
For example, Huang et al. \cite{huang_model_2014} applied an RC model to an airport terminal and demonstrated energy savings of 5–18\% via MPC. Additionally, improvements in thermal comfort are a crucial control scenario. For example, Arroyo et al. \cite{arroyo_identification_2020} achieved an 88.4\% reduction in total discomfort using an RC model, compared to the baseline BOPTEST controller \cite{Boptest}.

To enable RC-model-based MPC, suitable RC parameters must first be identified, usually by minimizing a least-squares error between measured and simulated temperatures. GAs are often used for this due to their robustness to non-convex loss landscapes and less sensitivity to initial parameter guesses \cite{arendt_modestpy_2019, SERASINGHE2024114123, BLUM2019410, EJENAKEVWE2025113368, GAO2019364}. Nonetheless, good initial parameter guesses can still improve estimation performance for GAs and speed up the estimation process \cite{SERASINGHE2024114123}. 
One way to mitigate this sensitivity is to manually define plausible parameter ranges based on physics and the literature, as noted in \cite{HARB2016199}. Similarly, Gao et al. \cite{GAO2019364} employed physical constraints and sampled initial parameter seeds randomly within those bounds for a GA-based optimization. Thereby, they achieved MAE values of 0.42--1.14 on a two-day test set across different RC model orders. Such strategies, however, depend on expert knowledge, which is costly to obtain and difficult to scale across buildings. 
Therefore, Blum et al. \cite{BLUM2019410} investigate the difference in a wide and a narrow parameter range for initialization. Wider ranges are easier to obtain, while narrow ranges can only result from an accurate analysis of the specific building. They reported RMSEs of 0.86 and 1.72 for the narrow and wide ranges, respectively, yielding a 20\% difference in MPC cost. 
When reliable physical ranges are unavailable, an alternative is to combine a global search algorithm with a subsequent local one. For instance, Li et al. \cite{LI2015139} used a GA to obtain an initial solution and then applied pattern-search for fine-tuning. Nevertheless, they reported a strong dependence on the global search result: if the GA estimate is too far from the global optimum, the local optimizer fails to converge to a satisfactory solution. Arendt et al. \cite{arendt_modestpy_2019} combined a rough physics-based initialization with a global search and a local one. They found GA plus sequential least squares programming to work best for their problem, though pure GA achieved comparable results at lower computational cost. 
Their case study is limited to a single-zone building based on an EnergyPlus simulation and a small dataset of 4 days (3 days train, 1 day test).
In summary, these studies typically evaluate only a small number of buildings and propose individual, non-generalizable strategies. Nonetheless, the literature often employs GAs for their strong global search capabilities. This robustness, however, comes at the cost of substantial computational overhead \cite{SERASINGHE2024114123, arendt_modestpy_2019, BLUM2019410}.

Beyond the choice of initial parameter ranges, the informativeness of the training data itself affects estimation quality \cite{SERASINGHE2024114123}: Bacher and Madsen \cite{bacher_identifying_2011} and Madsen and Schultz \cite{madsen1993short}, for example, show that using excitation signals, such as pseudo-random binary sequences, improves the identifiability of RC parameters compared to data collected under normal operation. However, such excitation is often infeasible in occupied buildings, as it can cause comfort violations and additional energy costs, motivating estimation methods that are robust to the limited excitation typical of operational data.

\section{Method}
\label{ch:method}

This section outlines our methodology. Section~\ref{ch:rc_circuits} introduces RC building models. Section~\ref{ch:nnestimation} presents our neural parameter estimation approach, the Estimator from Scratch. Section~\ref{ch:gen_estim} extends this with a pretrained network, the Pretrained Estimator. In Section~\ref{ch_m_data} we introduce the dataset used for pretraining and testing. Finally, Section~\ref{ch:benchmark} describes the baseline methods.

\subsection{Parameter Estimation of RC Models}
\label{ch:rc_circuits}

Parameter estimation in general relies on the assumption that a physical system can be approximated by a set of differential equations (gray-box model) of the form
\begin{equation}
\frac{d\mathbf{x}}{dt} = f_{\boldsymbol{\theta}}(\mathbf{x}, \mathbf{u}, \mathbf{d}),
\label{eq:state_eq_general}
\end{equation}
where $f(\cdot)$ is known and $\boldsymbol{\theta}$ unknown. $\mathbf{x}$ represents the system states, $\mathbf{u}$ control inputs, and $\mathbf{d}$ external disturbances. Parameter estimation determines $\boldsymbol{\theta}$ such that trajectories of $f_{\boldsymbol{\theta}}(\cdot)$ best match the measurement data. For buildings, $\boldsymbol{\theta}$ denotes the vector of RC parameters and $f_{\boldsymbol{\theta}}(\cdot)$ the chosen RC circuit. We use the two most common RC configurations (see Fig.~\ref{fig:rc_figure}):

\begin{figure}[t]
    \vspace*{-0.5cm}
    \centering
    \includegraphics[width=0.85\linewidth, trim={0.9cm 0 1.2cm 0},clip]{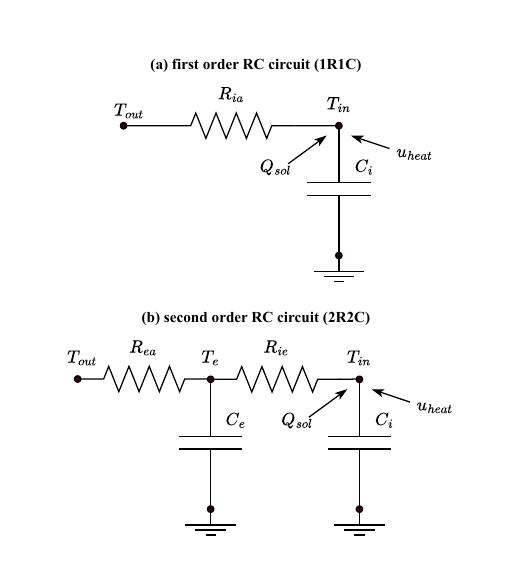}
    \vspace*{-0.6cm}
    \caption{Considered RC circuits defining ${d\mathbf{x}}/{dt} = f_{\boldsymbol{\theta}}(\mathbf{x}, \mathbf{u}, \mathbf{d})$.}
    \label{fig:rc_figure}
    \vspace*{-0.3cm}
\end{figure}

\textbf{1R1C model:} We first consider the widely used 1R1C model \cite{bacher_identifying_2011, huang_model_2014}, which is characterized by a single dominant time constant, low parameter count, and thus favorable identifiability. With indoor temperature $T_{in}$ as state $\mathbf{x}$, outdoor temperature $T_{out}$ and solar irradiation $Q_{solar}$ as disturbances $\mathbf{d}$, and heat source power $Q_{heat}$ as control signal $\mathbf{u}$, the 1R1C circuit \cite{bacher_identifying_2011, huang_model_2014} yields
\begin{align}
\frac{d T_{in}}{dt}
&= \frac{T_{out} - T_{in}}{R_{ia} C_i} + \frac{A_{eff}}{C_i}Q_{solar} + \frac{1}{C_i}Q_{heat}.
\label{eq:1r1c_ct}
\end{align}
Here, $R_{ia}$ is the interior--ambient resistance, $C_i$ the heat capacity, and $A_{eff}$ the effective area for solar gains. Although it is often referred to as RC parameter estimation, $A_{eff}$ is also included, resulting in three parameters to estimate.

\textbf{2R2C model:} To capture distinct indoor and envelope dynamics and more realistic heat exchange, the literature employs the 2R2C model \cite{EJENAKEVWE2025113368, arendt_modestpy_2019, bacher_identifying_2011, 7551202, bs2017_183}. This model adds a hidden envelope temperature $T_e$ and reads:
\begin{align}
\frac{d T_{in}}{dt}
&= \frac{T_e - T_{in}}{R_{ie} C_i} + \frac{A_{eff}}{C_i} Q_{solar} + \frac{1}{C_i} Q_{heat},
\label{eq:2r2c_ti}\\
\frac{d T_e}{dt}
&= \frac{T_{in} - T_e}{R_{ie} C_e} + \frac{T_{out} - T_e}{R_{ea} C_e}.
\label{eq:2r2c_te}
\end{align}
With the additional resistances $R_{ie}$ (interior–envelope) and $R_{ea}$ (envelope–ambient), and the capacity $C_e$ (envelope), this results in five parameters. The initial state $T_e$ is calculated via a voltage-divider approximation based on \cite{ROUCHIER201986, buildingenergygeeks}: $T_{e,0} = (R_{ie} T_{out,0} + R_{ea} T_{in,0}) / (R_{ie} + R_{ea})$.

\begin{figure*}[t]
    \centering
    \vspace*{-0.3cm}
    \includegraphics[width=1.05\linewidth]{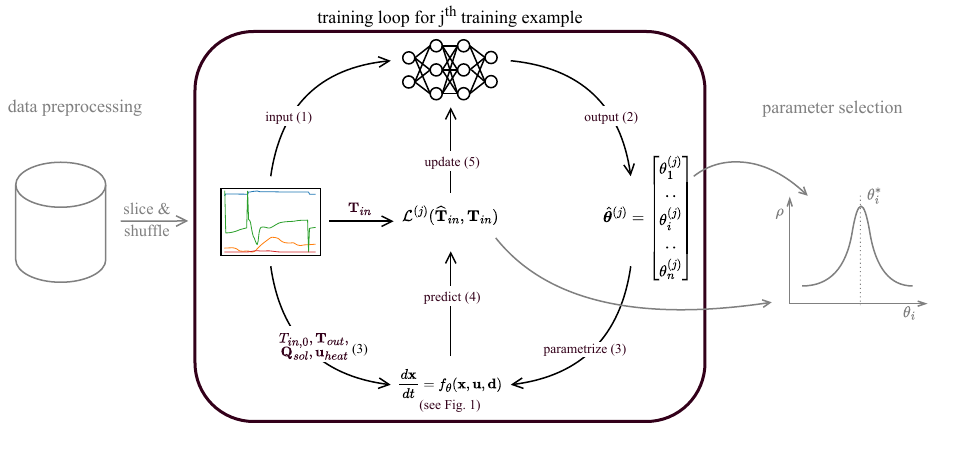}
    \vspace*{-1cm}
    \caption{Concept of neural parameter estimation for control-oriented systems. A neural network maps training examples sliced from measurement data to parameter estimates $\hat{\boldsymbol\theta}^j$. These estimates are used to parameterize a chosen RC model, which is then simulated to calculate the loss for the neural network update.}
    \label{fig:concept}
\end{figure*}

\subsection{Neural Parameter Estimation -- Estimator from Scratch}
\label{ch:nnestimation}

In this section, we introduce our neural parameter estimation approach for the estimation of the parameters $\boldsymbol{\theta}$. 

\textbf{General concept:} The general concept is illustrated in Fig.~\ref{fig:concept}. For data preprocessing, the available measurement time series data is sliced into training examples, similar to common practice in ML-based forecasting. Each training example contains trajectories of $\mathbf T_{in}$, $\mathbf Q_{heat}$, $\mathbf Q_{solar}$, and $\mathbf T_{out}$ of a fixed window length, the lookback. The examples are then shuffled and iteratively fed into the MLP-based training loop: in step (1), the neural network receives one training example as input; in step (2), it outputs a parameter vector estimate $\hat{\boldsymbol\theta}$, thereby mapping a time-series-based training example to constant parameters. Depending on the selected RC configuration, the output dimension is $n=3$ or $n=5$. To calculate the loss for updating the MLP's weights, a temperature forecast is required. To this end, the output estimate $\hat{\boldsymbol\theta}$ parametrizes the chosen RC network $f_{\hat{\boldsymbol{\theta}}}(\cdot)$ in step (3). This RC model is then simulated over a horizon of lookback$+1$ by a numerical solver, given an initial condition $T_{in,0}$ and the forcings $\mathbf Q_{heat}$, $\mathbf T_{out}$, $\mathbf Q_{solar}$ from the training example. This yields a predicted trajectory $\widehat{\mathbf T}_{in}$ in step (4), which is compared to the ground truth $ \mathbf T_{in}$ using the $\ell_2$ norm for the loss $\mathcal{L}$. The MLP parameters $\Phi$ are then updated in step (5) via $\nabla_\Phi \mathcal{L}$, by backpropagating through the solver. Differentiation is handled by the autodifferentiation scheme of PyTorch \cite{pytorch_autograd_2022}. The training loop (steps 1-5) is repeated for all training examples over 50 training epochs, with the examples reshuffled each epoch. 

Finally, we expect a single parameter estimate $\theta^\star$, which includes the vector of the final RC parameters. However, the neural network will produce different estimates across training examples, as each training example represents only a short time span and captures different values and dynamics. Moreover, individual optimal parameters (e.g. $R_{ia}^\star$ and $C_i^\star$) could arise from different training examples. We therefore follow \cite{doi:10.1073/pnas.2216415120}, who employ a marginalization step that weights the estimated parameters $\hat{\boldsymbol\theta}^j$ during the training process with their corresponding loss $\mathcal{L}^{(j)}$. This results in a histogram over plausible values for each individual parameter $\theta_i$, as illustrated in Fig.~\ref{fig:concept} (parameter selection). As the final parameter estimate, we select the value corresponding to the highest probability. Appendix~\ref{app:marginals} gives further detail on this step.

\textbf{Hyperparameters:} For training, we use the Adam optimizer \cite{kingma2014adam} and a batch size of 1. Tuning the remaining hyperparameters yielded little benefit across building datasets. Even when tuning on a dataset comprising multiple buildings (see Section~\ref{ch:gen_estim}), the same configuration emerged as for the single-building case. This behavior may stem from the non-standard nature of the learning task compared to typical ML applications, suggesting that a particular set of hyperparameters may be well-suited for parameter estimation of RC building models. Consequently, we selected a single hyperparameter configuration, detailed in Tab.~\ref{tab:hyperparameter}, that consistently achieved satisfactory performance.

\begin{table}[!b]
    \centering
    \begin{tabular}{p{1.2cm} p{1.1cm} p{0.8cm} p{1.3cm} p{0.9cm} p{0.9cm} }
        \hline 
        \textbf{Parameter} & Lookback & Layers & Neurons per layer & Output size & Input size \\ 
        \textbf{Selected}  & 96 & 2 & 140 & $n\times 1$ & $96 \times 4$ \\ 
        \hline
    \end{tabular}
    \caption{Hyperparameter selection for the MLP of the Estimator from Scratch and the Pretrained Estimator.}
    \label{tab:hyperparameter}
\end{table}

\textbf{Initialization:} In building parameter estimation, practitioners often specify plausible parameter ranges from which a random initial parameter guess is sampled, which improves convergence and overall performance \cite{SERASINGHE2024114123} (see Section~\ref{ch:literature}). We adopt the same prior knowledge and define parameter ranges based on \cite{HARB2016199}, which use a similar domain of buildings. We randomly draw a parameter seed from this range to initialize the MLP. To do so, the MLP is trained to predict the initial parameter seed as a constant output, given inputs uniformly sampled from $[0,1]$. This yields an initialized MLP network that can be applied to a target building for parameter estimation. Similar to other estimation algorithms, the estimation process can be repeated with multiple seeds to mitigate dependence on initial guesses and to enhance results \cite{SERASINGHE2024114123}. We select the number of seeds to 8, as a tradeoff between computational time and robustness. 
For each seed, we collect all parameter-estimate-loss pairs from the training loop to enrich the data for the marginalization step as explained in Appendix~\ref{app:marginals}.


\subsection{Pretrained Estimator}
\label{ch:gen_estim}

The Estimator from Scratch still relies on an initialization scheme based on an initial guess, which requires a plausible parameter range to draw from, multi-seed runs, and good initial guesses by chance. Moreover, the network is initialized by training on a static output and randomly sampled inputs, which do not represent building dynamics data. As a result, the network weights may be suboptimally initialized. 
To address these two limitations, we propose initializing the parameter estimation for a target building using a neural network pretrained on multiple source buildings, which we refer to as the \textit{Pretrained Estimator}. This idea builds on a TL paradigm, which uses multi-source pretraining to enhance modeling of a target building, as employed for black-box building models \cite{raisch2025gentlgeneraltransferlearning, koch2026thermalgems, fm_nagy, mpc_excite}.

The pretraining of the Pretrained Estimator relies on time-series data from 32 source buildings (data specifications will be provided in Section~\ref{ch_m_data}). This number follows Koch et al.~\cite{koch2026thermalgems}, who show that pretraining a fully black-box model on 32 source buildings is sufficient to outperform time series foundation models when applied to a target. 
For pretraining, we slice each of the source time series into training examples, as described in Section~\ref{ch:nnestimation}. Then we feed the training examples into the training loop shown in Fig.~\ref{fig:concept}, with two adjustments to enhance pretraining generalization: first, we shuffle the training examples across all source buildings rather than within a single building. Second, we use a larger batch size of 32 to incorporate information from multiple training examples and buildings into each neural network weight update. Training is performed for 30 epochs.
Once trained, we apply the Pretrained Estimator to an unseen target building via weight initialization \cite{pinto2022sharing}: the network's pre-learned weights serve as a starting point, from which a new Adam optimizer (not the one used for pretraining) seeks the specific minimum for the target data. This optimization then proceeds according to the general estimation process illustrated in Fig.~\ref{fig:concept}.

The Pretrained Estimator provides a generally pretrained neural network based on building dynamics data, rather than mapping uniformly sampled inputs to a static output. Thereby, it eliminates the need to draw an explicit initial parameter guess, since the initialization is instead encoded in the network weights themselves. As a result, it requires only a single seed rather than the multi-seed runs used for the Estimator from Scratch. 

\subsection{Source and target datasets}
\label{ch_m_data}

\begin{table}[!b]
    \caption{Distribution of physical properties for the source buildings, from~\cite{raisch2025gentlgeneraltransferlearning}: insulation U-value ($U_{wall}$), area-specific heat capacity ($c_{wall}$), window-to-wall ratio ($f_{win}$), and ground area ($A_{ground}$) with the remaining building properties calculated from these values \cite{krug2025highly}.}
    \label{tab:source_data}
    \vspace{-0.3cm}
     \centering
    \begin{tabular}{ccp{4cm}} \hline
         \textbf{Parameter}&  \textbf{Unit} & {\textbf{Sources}}  \\ \hline 
         $U_{wall}$ & $[W/m^2K]$ &  $\{0.1, 0.4, 0.7, 1, 1.3\}$ \\ 
         $c_{wall}$& $[kJ/m^2K]$ & $\{30, 165, 300\}$ \\ 
         $f_{win}$& - & $\{0.15, 0.20\}$  \\ 
         $A_{ground}$& $[m^2]$ & $\{60, 90, 120\}$ \\ 
         Weather & - & Belgrade, Prague, Berlin, London, Zurich \\ \hline
    \end{tabular}
\end{table}

For pretraining the Pretrained Estimator, we use source buildings, while both neural estimators are tested on target buildings. For the sources and targets, we use the simulated data from \cite{raisch2025gentlgeneraltransferlearning}, as they provide distinct datasets specifically designed for transfer learning use cases and have been applied in several studies \cite{RAISCH2026116868, mpc_excite, koch2026thermalgems}. The corresponding simulation environment is also provided, enabling closed-loop control assessment. Additionally, we evaluate real-world target buildings for prediction performance, as further explained in Section~\ref{ch:eval_pred}.

The source and target datasets from \cite{raisch2025gentlgeneraltransferlearning} stem from the BuilDa simulation framework \cite{builDaReview, krug2025highly}. This framework models single-zone buildings with different envelope components, sizes, locations, and occupancy profiles in Modelica and is validated according to ANSI/ASHRAE 140-2004 \cite{ASHRAE140}. \cite{raisch2025gentlgeneraltransferlearning} employ the distribution of single-family houses in Central Europe built between 1949 and today, according to \cite{tabula, TASK442013, Thomas2006Environmental}, while sources and targets differ in location, building properties, and occupancy. The source distribution of building properties is shown in Tab.~\ref{tab:source_data}. We sample a random subset of 32 sources from this distribution, as recommended by \cite{koch2026thermalgems}. For the targets, \cite{raisch2025gentlgeneraltransferlearning} provide a representative set of 8 buildings, as specified in Tab.~\ref{tab:8targets}, covering a predefined target distribution. 
Each building time series covers 3 months (January to March) of winter data, the period when the heating signal $Q_{heat}$ is active, sampled at 15-minute intervals.


\newcommand{\rot}[1]{\multicolumn{1}{c}{\hspace{-1ex}\adjustbox{angle=30,lap=\width-1em}{#1}}}
\begin{table}[!b]
    \caption{Properties of target buildings \cite{raisch2025gentlgeneraltransferlearning} with the daytime setpoint $T_{sp,\,day}$ and a potential night setback $\Delta T_{night}$.}
     \label{tab:8targets}
     \vspace{-0.4cm}
     \centering
    \begin{tabular}{c@{\hskip .33cm}|ccccccc} 
        \rot{Target building} & \rot{$U_{wall}$ $[W/(m^2K)]$} & \rot{\textbf{$c_{wall}$ $[kJ/(m^2K)]$}} & \rot{\textbf{$f_{win}$}}  & \rot{\textbf{$A_{ground}$ $[m^2]$}} & \rot{$T_{sp,\,day}$ [\textdegree C] } & \rot{$\Delta T_{night}$ [\textdegree C]} & \rot{Weather} \\  \hline
        T1 & 0.25 & 280 & 0.19 & 100 & 21.0 & 2.0 & Amsterdam \\  
        T2 & 0.25 & 40  & 0.16 & 70 & 22.0 & 1.0 & Bratislava \\
        T3 & 0.55 & 150 & 0.16 & 70& 23.0 & 3.0 & Amsterdam \\
        T4 & 0.55 & 280 & 0.19 & 100 & 20.5 & 1.5 & Munich \\
        T5 & 0.85 & 150 & 0.19 & 100 & 22.5 & 0.5 & Bratislava \\
        T6 & 0.85 & 40 & 0.16 & 70 & 22.0 & 2.5 & Munich \\  
        T7 & 1.15 & 280 & 0.16 & 70 & 23.0 & 0.0  & Bratislava \\ 
        T8 & 1.15 & 40 & 0.19 & 100 & 23.0 & 1.5 & Amsterdam \\ \hline
     \end{tabular}
\end{table}

\subsection{Benchmark Methods}
\label{ch:benchmark}

For benchmarking the neural estimators, we use both a conventional RC estimator and a fully black-box baseline.

\textbf{Genetic Algorithm (GA):} For the conventional RC estimator, we use a GA, as it is most commonly used for RC parameter estimation in buildings \cite{arendt_modestpy_2019, SERASINGHE2024114123, BLUM2019410}. The GA is a gradient-free optimization method inspired by natural evolution that minimizes the loss (MSE) between measured and simulated trajectories by evolving populations of potential solutions through selection, crossover, and mutation. GAs offer robustness to non-convex landscapes and are less sensitive to imperfect or missing initial parameter guesses, making them especially favorable for buildings \cite{SERASINGHE2024114123}. 
However, to improve performance, we apply the same initial parameter range and number of seeds as done for the Estimator from Scratch (see Section~\ref{ch:nnestimation}). Among all seeds, the estimate with the lowest training loss is selected.
For the implementation, we use the GA algorithm and hyperparameters from \cite{arendt_modestpy_2019}, who provide an open-source implementation for building RC parameter estimation.
We also evaluated a gradient-based least-squares approach using the Adam optimizer. This estimation did not yield improvements over the GA method, and the corresponding results are therefore omitted.

\textbf{MLP benchmark:} To test whether the added complexity introduced by neural parameter estimation is justified, we also test its fully black-box alternative. Accordingly, we use an MLP benchmark that directly maps $f(\cdot)$ (Eq.~\eqref{eq:state_eq_general}), rather than estimating $\boldsymbol\theta$ of a predefined gray-box structure. Since Stoffel et al.~\cite{stoffel2023evaluation} found an MLP model to perform best among several white-, gray-, and black-box models, we reuse their MLP implementation and hyperparameters: 2 hidden layers with 33 neurons each. Thus, the MLP benchmark has fewer neurons per layer than the neural estimators. This is done intentionally, as smaller networks are often better in downstream tasks, potentially due to better extrapolation capacities \cite{mpc_excite, stoffel_real-life_2024, stoffel2023evaluation}. The training procedure of the MLP benchmark closely mirrors that of the Estimator from Scratch, except that the MLP benchmark outputs an indoor temperature forecast over the entire horizon directly (24 h, cf. Section~\ref{ch:eval_pred}), without the physics-informed RC structure in between. To prevent overfitting, we hold out the last 20\% of the training set as a validation set and select the best model based on validation performance across training epochs.

\section{Evaluation}
\label{sec_setup}

For evaluation, we follow the standard procedure for RC-based modeling in buildings \cite{BLUM2019410, huang_model_2014, bs2017_183, LI2015139, arroyo_identification_2020}: we assess prediction performance on a held-out test set, and closed-loop control performance via MPC, which is the ultimate purpose of RC parameter estimation. Both are detailed below. Section~\ref{ch:eval_compute} additionally discusses computational time.

The code for the neural estimators, the benchmarks, and the experiments is available at \cite{GitHub}.

\subsection{Prediction Performance}
\label{ch:eval_pred}

\begin{figure}[t]
    \centering
    \includegraphics[width=1\linewidth, trim={0.4cm 0 0.4cm 0},clip]{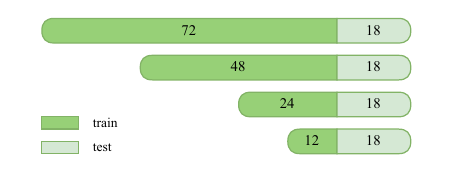}
    \vspace{-1.1cm}
    \caption{Setup for variable training lengths. The test set remains fixed.}
    \label{fig:train_n_test}
    \vspace{-0.3cm}
\end{figure}

For prediction performance, we compare the models' trajectories against the ground truth over a 24-hour horizon (96 steps), as in \cite{arendt_modestpy_2019, 7551202, DECONINCK2016290}. This comparison is performed on each target building's test set, which follows the training set in consecutive order (Fig.~\ref{fig:train_n_test}). We use the root mean square error (RMSE) as the primary evaluation metric. For comparisons across buildings with different temperature ranges, we additionally report the normalized RMSE (NRMSE), normalized by the range (max. minus min.) of indoor temperature in each building's test set. The mean absolute error was also evaluated but showed the same pattern as RMSE and is therefore omitted due to limited space. To compare against a benchmark method, we report the relative improvement in RMSE\footnote{(RMSE$_{benchmark}$ $-$ RMSE$_{model}$)$/$RMSE$_{benchmark}$.}. In practice, the amount of data available for estimation depends on the sensor installation date, which in turn determines how early energy-efficient control systems can be deployed. Moreover, \cite{BLUM2019410} found a reasonable influence of training data length on test set performance. We therefore evaluate four training data lengths for the simulated buildings (12, 24, 48, 72 days), with the test set fixed to the same period immediately following training (see Fig.~\ref{fig:train_n_test}). For the real-world data, detailed below, we use three data lengths (12, 24, 48 days) due to the limited data availability.

\textbf{Real-world data:} For the prediction assessment, we additionally use a set of three real-world buildings. Real-world data suitable for RC modeling are scarce because $Q_{heat}$ is rarely measured directly in practice. First, we use the \textit{Varennes Net-Zero Energy library} \cite{SARTORI2023109149}. This dataset comprises data collected over a 2-month period. The building is a two-story library with a floor area of about 2100 m$^2$, located near Montreal, Canada. Compared to the simulated data (single-family houses in Central Europe), this dataset displays a non-residential building in North America.
Second, we use the two real-world \textit{ThermBuild TwinHouses} (TwH1, TwH2) \cite{raisch2026thermbuild}, located at the Fraunhofer IBP campus in Holzkirchen, Germany. Both buildings represent single-family houses in Central Europe and exhibit envelope characteristics typical of an average newly refurbished German single-family house. The buildings differ in their occupancy behavior and their heating system: TwH1 uses a compact heat pump with a small 50\,L buffer (direct-coupled), while TwH2 uses a heat pump with a separate 825\,L domestic hot water tank and a 500\,L heating buffer (storage-coupled). 
All real-world data is provided at 15-minute resolution.

\subsection{Model Predictive Control}
\label{ch:eval_mpc}

Next, we assess whether the models' prediction performance translates into closed-loop control performance. To do so, we follow the formulation of \cite{stoffel_real-life_2024, STOFFEL2023100296, mpc_excite}:
\begin{subequations}
\label{eq_mpc}
\begin{align}
\min_{\mathbf{u}, \mathbf{\epsilon}} \quad
& \sum_{k=0}^{H_p-1} \Big[ w_0\epsilon_k^2 + w_1 \, u_k^2 + w_2(u_k - u_{k-1})^2 \Big] \label{eq_mpc_obj}\\
\text{s.t.} \quad
& T_{k+1} = f_\theta\!\left(T_{k},\, d_{k},\, u_{k}\right), \label{eq_mpc_model}\\
& T_k^{\text{low}} - \epsilon_k \leq T_k \leq T_k^{\text{up}} + \epsilon_k, \label{eq_mpc_comfort}\\
& \epsilon_k \geq 0, \qquad 0 \le u_{k} \le 1, \label{eq_mpc_slack}\\
&\forall k \in [0, \dots, H_p-1], \label{eq_mpc_k}
\end{align}
\end{subequations}
where $k$ indexes the prediction horizon $H_p$, $f_\theta(\cdot)$ is the estimated model (RC model or MLP benchmark), $T_k$ is the indoor temperature, $d_k$ the weather disturbances, and $u_k$ the heating control signal. The objective penalizes comfort-band violations via $w_0$ and the slack variable $\epsilon_k$, with time-varying bounds $T_k^{\text{low}}$, $T_k^{\text{up}}$ set as in Tab.~\ref{tab_Tset}, similar to the BOPTEST convention \cite{Boptest, counterdyna}. A smoothing term ($w_2$) avoids control oscillations, and an energy term ($w_1$) penalizes heating use. The cost function parameters in Eq.~\eqref{eq_mpc_obj} are set to $w_0 = 100$, $w_1 = 1$, and $w_2 = 4$, based on \cite{stoffel2023evaluation, raissi2019physics}. The comfort weight $w_0$ is set high, since comfort violations should not be traded off against energy savings. 
We note that, as with any multi-objective optimization, weight selection remains inherently heuristic.
We omit an explicit electricity-price objective since it would introduce an additional degree of freedom: the choice of price signal. 

At each control step, the MPC controller selects a trajectory $u$ that minimizes Eq.~\eqref{eq_mpc_obj} based on the selected model $f_\theta(\cdot)$. The first control action of the trajectory is then applied to the target building, the resulting building response is observed, and the process repeats in a receding-horizon fashion.
The optimization in Eq.~\eqref{eq_mpc} is solved depending on the model type. For the RC models, we rewrite the 2R2C model as a state-space model similar to \cite{bs2017_183}, yielding a convex quadratic program that can be solved efficiently during MPC. The MLP benchmark defines a non-convex but differentiable mapping, which we instead optimize via gradient descent. Gradients of the cost with respect to the control sequence are obtained via automatic differentiation in PyTorch \cite{paszke2019pytorch} and run for 200 iterations as done in \cite{mpc_excite}.

\begin{table}[tb]
    \caption{Comfort range for the indoor temperature.}
    \label{tab_Tset}
    \vspace{-0.3cm}
    \centering
    \begin{tabular}{p{3.7cm}|p{1.5cm}|p{1.5cm}}
    Time & $T^{\text{low}}$  & $T^{\text{up}}$ \\ \hline
    Mon.--Fri. 6:00--18:00 & 17\,\si{\celsius} & 24\,\si{\celsius} \\
    Mon.--Fri. 18:00--6:00 & 20\,\si{\celsius}& 21\,\si{\celsius}\\
    Sat.--Sun. & 20\,\si{\celsius}& 21\,\si{\celsius}\\
    \end{tabular}
    \vspace{-0.4cm}
\end{table}

We evaluate MPC on the 8 simulated buildings, since closed-loop control requires a live building model to interact with, which is not available for the real-world datasets. Control testing is performed on the same period as the prediction assessment.

As an additional baseline, we use BuilDa's integrated proportional-integral (PI) controller \cite{krug2025highly}, a standard, model-free approach common in real buildings. It regulates the heating signal $u$ from the error between a setpoint $T^{\text{set,PI}}$ and the measured indoor temperature. We set $T^{\text{set,PI}}=T^{\text{low}}$, the lower bound of the comfort band (Tab.~\ref{tab_Tset}), to minimize heating while meeting comfort requirements.

\subsection{Computational Time}
\label{ch:eval_compute}

Experiments were conducted on an Intel Xeon Platinum 8480+ CPU with 112 cores and 488\,GiB of RAM. The Estimator from Scratch required, on average, 12 min to estimate for one seed. The Pretrained Estimator was fine-tuned for the same number of epochs, resulting in the same per-target computational cost but faster convergence and requiring only one seed. Pretraining across the 32 source buildings took 18 min. The GA remained the most expensive method, requiring an average of 5 h and 18 min per seed\footnote{Runtime could be improved by parallelizing the GA's selection, crossover, and mutation steps.}. The multi-seed runs were executed in parallel across multiple cores.
The average MPC update time was 0.064\,s for the RC-based models, regardless of estimation method, and 0.976\,s for the MLP benchmark.

\section{Experiments}
\label{ch:experiments}

This section reports our results in two parts. Section~\ref{ch:exp_pred} evaluates prediction performance across simulated and real buildings, doing this for variable training data length and the two RC configurations. Section~\ref{ch:exp_mpc} then evaluates the closed-loop control performance on the simulated buildings.

\subsection{Prediction Performance}
\label{ch:exp_pred}

\begin{figure}[t]
    \centering
    \includegraphics[width=1\linewidth]{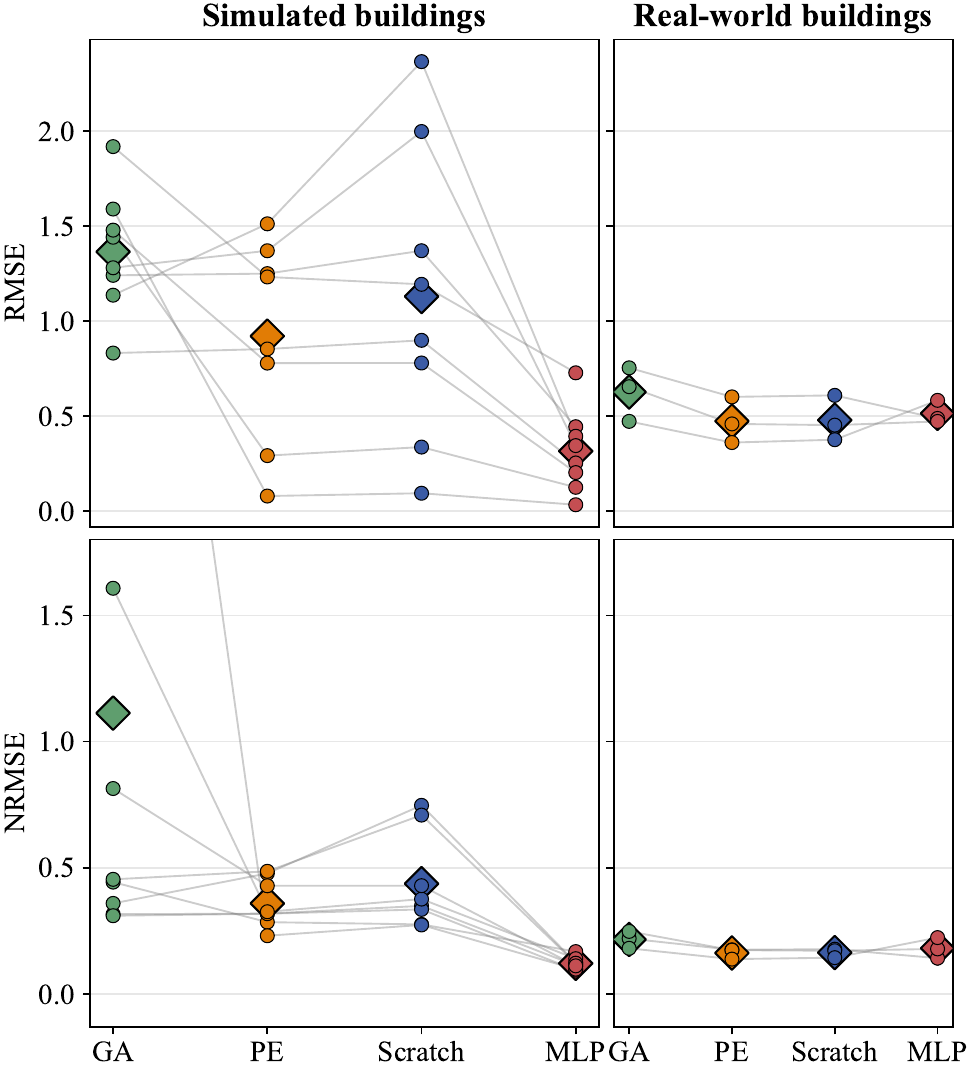}
    \vspace{-0.7cm}
    \caption{RMSE and NRMSE (in K) for the 2R2C model with 12 days of training data, shown as paired scatter plots for the simulated and real-world buildings. For GA, the NRMSE value of 4.60 is omitted for visual clarity.}
    \label{fig:pred_2r2c_12d}
    \vspace{-0.4cm}
\end{figure}

We first compare all four methods, the GA, the Estimator from Scratch (Scratch), the Pretrained Estimator (PE), and the MLP benchmark, on the 2R2C model with 12 days of training data, which is the most data-scarce scenario. Fig.~\ref{fig:pred_2r2c_12d} shows RMSE and NRMSE values averaged and as scatter for the simulated buildings and the three real buildings. For the simulated buildings, the Estimator from Scratch achieves an average RMSE of $1.13\,\si{\kelvin}$ and the Pretrained Estimator $0.92 \,\si{\kelvin}$, leading to an average improvement over the GA ($1.36 \,\si{\kelvin}$) of 16.9\% and 32.3\%, respectively. However, the Estimator from Scratch yields the largest range across buildings, which can be lowered through the Pretrained Estimator. For the Pretrained Estimator, we mostly see lower or equal errors compared to the GA-benchmark, except for one building. The MLP benchmark achieves the lowest range and average RMSE, at $0.32 \,\si{\kelvin}$, which is around 65--76\% lower than any RC-based method. However, the MLP provides no physical structure, which we show to be challenging for MPC in Section~\ref{ch:exp_mpc}. 
For NRMSE values, the GA benchmark has the largest range across buildings. This is due to larger prediction errors for buildings with small temperature ranges, in particular T7 and T5 (see Tab.~\ref{tab:8targets}). In contrast, the Pretrained Estimator exhibits a relatively low range, showing consistent results across buildings based on their temperature deviations.

For the real buildings, the ranking of methods is broadly consistent with the simulation results, though errors (except for the MLP, which now performs comparatively worse) are lower and less variable overall. This may be explained by the night setback present in the simulated buildings but absent in the real ones, resulting in more variable temperature trajectories in the simulated buildings. Averaged across the real buildings, the Pretrained Estimator achieves an RMSE of 0.47\,K, the Estimator from Scratch 0.48\,K, and the GA 0.62\,K, corresponding to a 24.19\% improvement of the Pretrained Estimator over the GA. Notably, the Pretrained Estimator remains effective on the real buildings despite being pretrained on simulated data, though its advantage over the Estimator from Scratch is now only marginal.
\\
\begin{figure*}[t]
    \centering
    \includegraphics[width=1\linewidth]{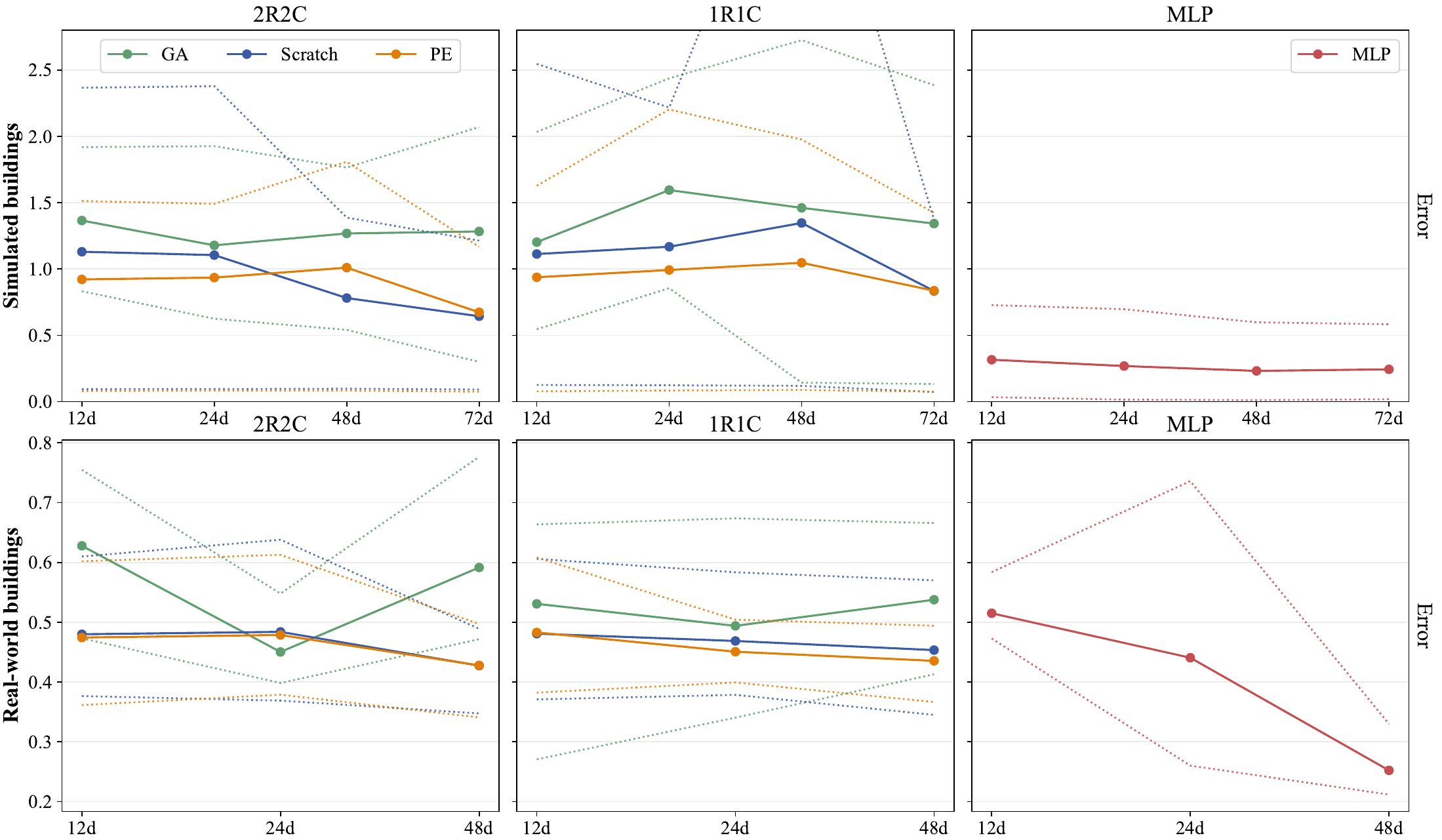}
    \vspace{-0.3cm}
    \caption{Average (full color) and min-max (dashed, light color) RMSE values in K for simulated and real-world buildings as a function of training data length for the 2R2C, 1R1C, and the MLP benchmark.}
    \label{fig:pred_length_sim}
\end{figure*}

\noindent
Next, we investigate the effects of the RC model structure (2R2C and 1R1C) and the training data length. Fig.~\ref{fig:pred_length_sim} reports the mean and min-max range across buildings for the simulated and real-world buildings. We omit individual building traces here to avoid clutter and provide them extra in Appendix~\ref{app:indu}. There, we show that the neural-based approaches outperform the GA-based estimation for 20 of 22 building-RC configuration combinations, with up to 72 days of training data available to each method.

Fig.~\ref{fig:pred_length_sim} shows the resulting average prediction error as a function of training data length for both RC configurations. For the simulated buildings, the Pretrained Estimator performs best at small data lengths, while the Estimator from Scratch catches up as more training data becomes available. Both methods improve with more data and converge to values of around 0.65\,K (2R2C) and 0.83\,K (1R1C) by 72 days. This indicates that pretraining is particularly beneficial when data is scarce. 
The GA-based estimation, by contrast, performs best at short training lengths (12-24 days), reaching an average RMSE of 1.18\,K (2R2C, 24 days) and 1.20\,K (1R1C, 12 days), but does not improve with additional data. Across both RC configurations, the 2R2C model achieves slightly better performance than the 1R1C model. The MLP's RMSE decreases slightly with more training data and remains far below all RC-based methods across all training lengths, though this does not hold for control, as we show later. 

For the real-world data, the Pretrained Estimator still outperforms the Estimator from Scratch, but now with a smaller margin. The GA performs worse compared to both neural approaches across most data lengths, except at 24 days for the 2R2C model. In contrast to the simulated buildings, the MLP benchmark achieves the lowest RMSE only at 48 days, whereas at 12 and 24 days the RC models perform better. This is likely because the MLP requires more data to compensate for the added noise and disturbances present in real-world measurements. 


\subsection{Control Performance}
\label{ch:exp_mpc}

Next, we evaluate all methods in closed-loop MPC on the 8 simulated buildings, as explained in Section~\ref{ch:eval_mpc}. We use the 2R2C model for the estimation methods as it performed slightly better than the 1R1C model. For each simulated building, we use its best prediction model in terms of training length for each method. This is done, as the GA often performed better for short and intermediate training lengths. 

\begin{figure}[t]
    \centering
    \includegraphics[width=1\linewidth]{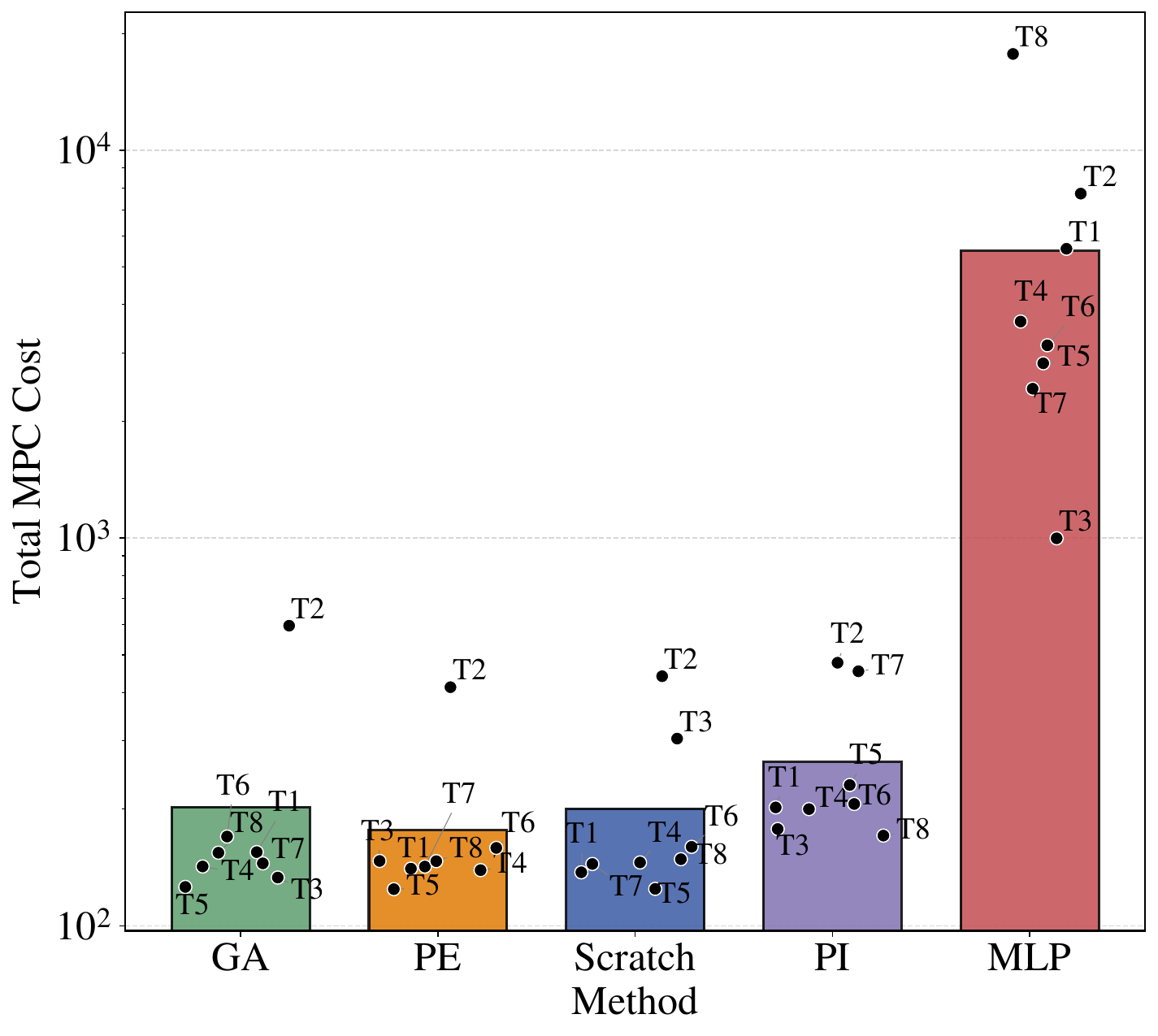}
    \vspace{-0.9cm}
    \caption{Total MPC cost (logarithmic scale) for all methods averaged (bar) and shown as scatter per building (T1--T8).}
    \label{fig:mpc_total_cost}
    \vspace{-0.5cm}
\end{figure}

Fig.~\ref{fig:mpc_total_cost} shows the total MPC cost averaged and as a scatter per building for the GA, Pretrained Estimator, Estimator from Scratch, the MLP benchmark, and the PI controller, with individual results in Tab.~\ref{tab:building-costs}. Averaged across all eight buildings, the Pretrained Estimator achieves the lowest mean total cost (176.51), resulting in a 12.87\% improvement over the GA (202.58), a 12.05\% improvement over the Estimator from Scratch (200.69), and a 33.36\% improvement over the PI controller (264.88). The Estimator from Scratch thus performs only slightly better than the GA, despite its lower prediction accuracy. 
The Pretrained Estimator also shows the lowest variations across buildings, with a minimum-to-maximum cost difference of 288.3, followed by the Estimator from Scratch (315.9), while the GA shows the highest range (468.4). The MLP, despite its strong prediction accuracy, achieves a mean total cost of 5502.63, indicating poor extrapolation capabilities compared to the RC-based models.

\begin{table}[b]
\centering
\caption{Total cost by method and building. Best (lowest) total cost per building in bold, second-best underlined.}
\label{tab:building-costs}
\vspace{-0.35cm}
\resizebox{\columnwidth}{!}{%
\begin{tabular}{@{}l rrrrrrrr@{}}
\toprule
Method & T1 & T2 & T3 & T4 & T5 & T6 & T7 & T8 \\
\midrule
GA      & 155 & 594 & \textbf{133} & \underline{142} & 126 & 170 & 145 & 154 \\
PE      & \underline{141} & \textbf{413} & \underline{147} & \textbf{139} & \textbf{124} & \textbf{159} & \textbf{142} & \textbf{147} \\
Scratch & \textbf{138} & \underline{440} & 304 & 146 & \underline{125} & \underline{160} & \underline{144} & \underline{149} \\
PI contr.     & 202 & 477 & 178 & 200 & 231 & 206 & 454 & 171 \\
MLP     & 5571 & 7729 & 999 & 3617 & 2823 & 3141 & 2428 & 17714 \\
\bottomrule
\end{tabular}%
}
\end{table}

According to Tab.~\ref{tab:building-costs}, the Pretrained Estimator achieves the best performance in 6 of 8 buildings and is second-best in the remaining two. The Estimator from Scratch is best in 1 of 8 buildings and second-best in 5, while GA is best in 1 building and second-best in 1. For building control results, shown for each MPC objective separately, we refer to Appendix~\ref{indu_control}.
\\
\begin{figure}[t]
    \centering
    \includegraphics[width=1\linewidth]{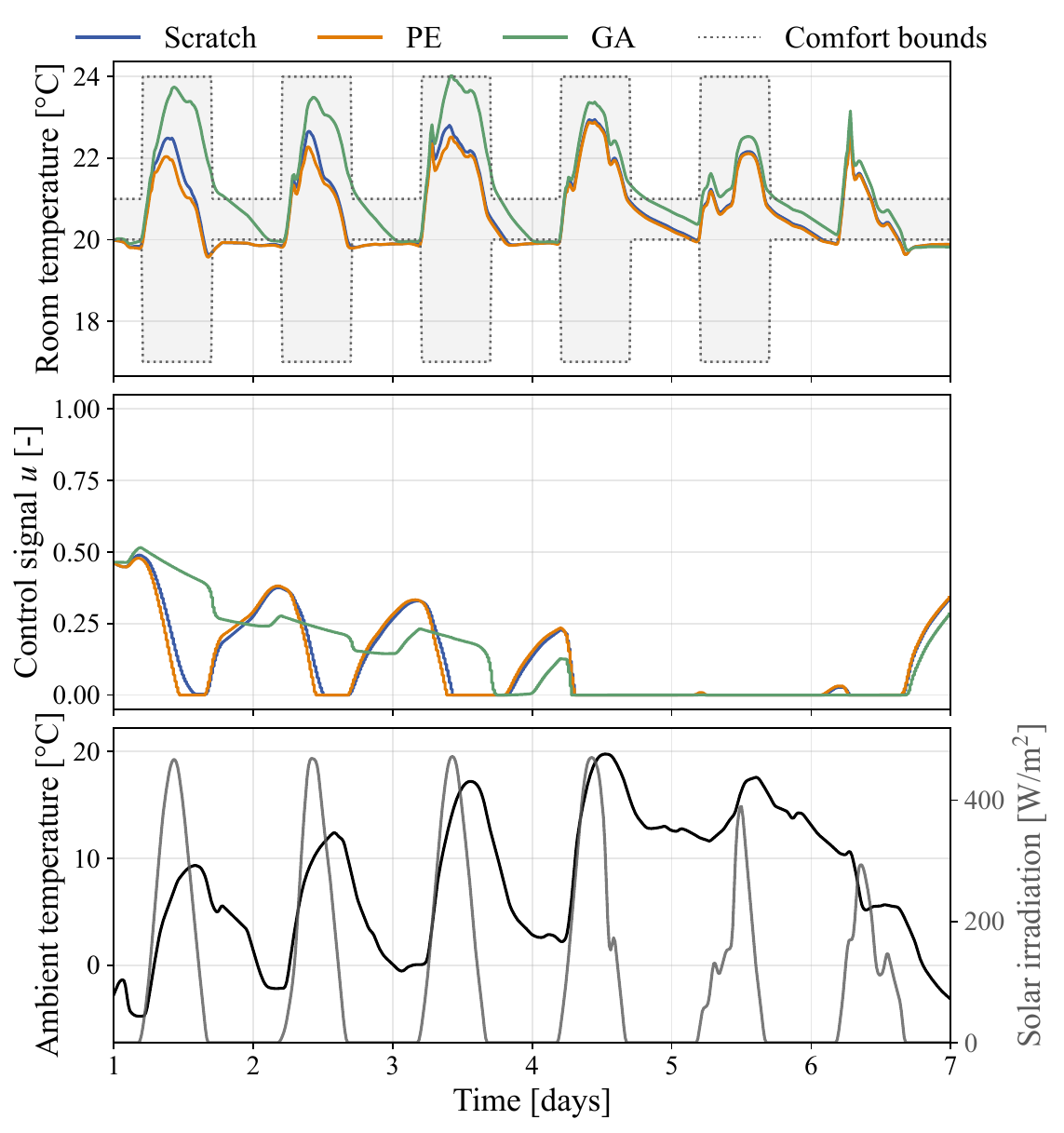}
    \vspace{-0.9cm}
    \caption{Resulting indoor temperature and control signal for building T2 for the first days of the MPC test period, with ambient temperature and solar irradiation shown as disturbances.}
    \label{fig:mpc_timeseries}
\end{figure}

\noindent
As an example, we plot the temperature and control trajectory for building T2 over the first days of the testing period in Fig.~\ref{fig:mpc_timeseries}. We use T2 to illustrate its outlier behavior observed in Fig.~\ref{fig:mpc_total_cost}. We only plot the RC-based controllers to reduce visual clutter. Fig.~\ref{fig:mpc_timeseries} illustrates that all three controllers keep the room temperature within the comfort bounds most of the time and modulate the control signal to track daily setpoint changes. The GA-based controller, however, uses more energy and reaches higher temperatures, likely because its prediction model is less accurate (see Appendix~\ref{app:indu}). During the day, solar irradiation mostly drives the indoor temperature, so the controller remains off. On the weekend (days 6--7), solar gains even exceed the comfort constraints, resulting in comfort costs that the controller cannot counteract. The combination of sunny Bratislava weather and T2's good insulation makes this building especially prone to upper comfort bound violations. However, since this holds across all controllers similarly, it does not affect the cost comparison.

\section{Discussion}
\label{ch:discussion}

This work introduced a neural parameter estimation framework for building RC models called the Estimator from Scratch. The Pretrained Estimator extends this by pretraining the network across multiple source buildings. We compared both methods against a GA-based RC estimation and a black-box MLP benchmark, across eight simulated and three real-world buildings, two RC configurations, and four training data lengths, along two paradigms: prediction accuracy and closed-loop MPC performance. Following, we summarize the main conclusions of this work:

\textbf{Prediction Performance.} We observe that GA-based estimation performs well for certain training data lengths but poorly for others, indicating a sensitivity to training data length similar to \cite{BLUM2019410}. This variable performance complicates its application, since multiple training data lengths must be investigated to ensure good results. In contrast, both neural estimators vary less with training data length and mostly improve with more data that becomes available. This robustness is also useful beyond a single estimation run: in a continuous learning setup, the Pretrained Estimator could serve as the initial starting point that is incrementally fine-tuned as more data becomes available, similar to \cite{RAISCH2026116868}.
Across 22 building-RC configuration pairs and up to 72 days of training data, the neural approaches achieve the best results in 20, mostly outperforming the GA on both simulated and real-world buildings. For the shortest training data scenario (12 days), the Pretrained Estimator outperformed the GA benchmark by 32.3\% on the simulated and 24.19\% on the real-world data. The Pretrained Estimator remains effective on the real buildings despite the SIM2REAL gap, though its edge over the Estimator from Scratch is now only marginal. Pretraining on real-world rather than solely simulated data may therefore be a promising direction to further enhance performance.

\textbf{Control Performance.} The Pretrained Estimator achieves the lowest mean and least variable MPC cost across buildings, outperforming the GA-based controller on average by 12.9\% and the PI controller by 33.4\%. The Estimator from Scratch and the GA achieve roughly the same control performance, despite the Estimator from Scratch's better prediction accuracy. This is a known phenomenon: higher prediction accuracy does not necessarily translate into better control performance \cite{BLUM2019410, mpc_excite, martin_poster}. This is because operational test data provides little information content and can only partially indicate control performance, as shown in \cite{mpc_excite}. The same phenomenon emerges for the MLP benchmark: it achieves the best prediction performance of all methods, yet also the highest MPC cost, likely because it extrapolates poorly beyond the operating conditions seen during training. As shown in \cite{mpc_excite, stoffel2023evaluation}, MLPs can reach good control performance, but only when trained on excitation-based data with broad state-action space coverage, which typical operational data does not provide. This underscores that RC models benefit especially from their predefined structure for control. 

\textbf{Neural Parameter Estimation.} This work brings neural parameter estimation to control-oriented dynamical systems and applies it to RC building models. In general, neural parameter estimation achieves comparable or improved prediction and control performance relative to the established GA benchmark, while offering several additional benefits. First, the neural parameter estimation approaches require significantly less optimization time than the GA, whose high computational burden has been repeatedly flagged in the literature \cite{SERASINGHE2024114123, arendt_modestpy_2019, BLUM2019410}. Second, neural-based approaches show a straightforward dependence on the amount of training data: more data generally improves performance, making them easier to apply than the GA, for which multiple training data lengths should be investigated to ensure good performance. Third, this work brings machine learning (ML) to RC parameter estimation, opening up new research directions in which recent advances from the ML community could be applied to RC modeling. We demonstrate this for transfer learning with the Pretrained Estimator. Further ideas could follow, such as meta-learning \cite{finn2017model}, prior-fitted networks \cite{muller2022transformers, koch2025_tpfn}, or foundation-model-style pretraining \cite{bommasani2021opportunities} to enhance parameter estimation.

Mechanistically, neural parameter estimation traverses the loss landscape using randomly sampled training examples. This implicitly assumes that the training data contains at least some informative windows, allowing the method to yield good estimation results even at non-excited data. While exploring the loss landscape with the training examples, the network weights are adjusted. This narrows down the plausible parameter space over time, which can be interpreted as a global search that gradually evolves into a local one. When using the Pretrained Estimator, the plausible parameter space is even initialized using knowledge from a pretraining phase.


\textbf{Pretrained Estimator.} The Pretrained Estimator shows one of the ML advances directly applied in this paper, i.e., transfer learning via pretraining on multiple source buildings. Thereby, the benefits from neural parameter estimation can be maintained while adding further ones: First, the embedded knowledge further improves prediction accuracy and control performance among target buildings. Second, the Pretrained Estimator eliminates the need to provide an explicit initial parameter guess, thereby requiring only a single-seed run per building. This makes estimation more robust and independent of how well an initial guess is drawn. However, because this knowledge is encoded in the network weights, it cannot be extracted to initialize conventional optimization-based methods.


\textbf{Limitations.} As this is the first neural parameter estimation approach for RC building models, this study naturally has limitations. First, we consider only a single neural network design. We also evaluated an LSTM architecture; however, it did not improve results, possibly due to the non-standard nature of the learning task, mapping time series to static parameters rather than to a forecast. 
Second, several parameter estimation studies use excitation-based data \cite{bacher_identifying_2011, madsen1993short}, such as pseudo-random binary sequences, to improve estimation results, which is an interesting direction for future evaluation of our approach. However, such excitation is often infeasible in real, occupied buildings, since it can cause comfort violations and additional cost \cite{SERASINGHE2024114123, Drgona.2020}. Accordingly, an estimation method performing well on unexcited data is preferred. Third, we used outdoor temperature and solar irradiation in the RC model only, as they are most commonly used and readily available in real-world scenarios. However, these models could be improved by using more sophisticated disturbance models, for example, for occupancy and ventilation, as in \cite{hu_building_2016, kim2025hybrid}. Finally, our control evaluation is limited to simulated buildings, since closed-loop MPC on the real-world buildings would require live access to them. Extending this evaluation to real-world closed-loop control remains an important next step.

\section{Conclusion}
\label{ch:conclusion}

We introduced neural parameter estimation for building RC models, which we call the Estimator from Scratch. We further extend this with a transfer learning approach that exploits knowledge across multiple source buildings, called the Pretrained Estimator. The Estimator from Scratch achieves competitive prediction and control performance relative to the Genetic Algorithm-based RC estimation benchmark, while being less sensitive to training data amount and requiring less computational time. The Pretrained Estimator maintains these benefits while additionally removing the dependence on an initial guess and outperforming the GA-based controller on average by 12.9\% and the PI controller by 33.4\% in closed-loop control. These results propose the Pretrained Estimator as a robust, computationally efficient, and initial-guess-free alternative to conventional RC parameter estimation, with potential for broader application to control-oriented dynamical systems.

\balance
\bibliographystyle{ACM-Reference-Format}
\bibliography{sample-base}


\appendix
\section{Marginal densities}
\label{app:marginals}

During training of the neural network, the space $\mathbb{R}^{n}$ of plausible parameters $\boldsymbol\theta$ is explored, where $n = \dim(\boldsymbol\theta)$ is the number of parameters to be estimated. Each training iteration $j$ yields a parameter estimate $\hat{\boldsymbol{\theta}}^{(j)}$ and its corresponding loss $\mathcal{L}^{(j)}$. This relation is used to calculate the marginals for each parameter:
\begin{equation}
    \rho (\theta_i) \propto \int \exp\big(-\mathcal{L}) d\theta_{-i}.
    \label{eq:marginals}
\end{equation}
Here $\theta_i$ denotes an element of the vector $\boldsymbol\theta$ corresponding to an R, C, or A parameter. Further, $d\theta_{-i}$ denotes an integration over all parameters $\boldsymbol\theta$ except parameter $\theta_{i}$. 
This converts the parameter-estimate-loss pairs seen during training into a histogram for each parameter $\theta_i$ individually (Fig.~\ref{fig:concept}, parameter selection). For each final parameter $\theta^\star_i$, we select $\theta^\star_i = \arg\max_{\theta} \rho(\theta_i)$ corresponding to the highest probability. For further explanation, we refer to \cite{doi:10.1073/pnas.2216415120}.


\section{Individual prediction results}
\label{app:indu}

\begin{figure*}[]
    \centering
    \includegraphics[width=0.7\linewidth]{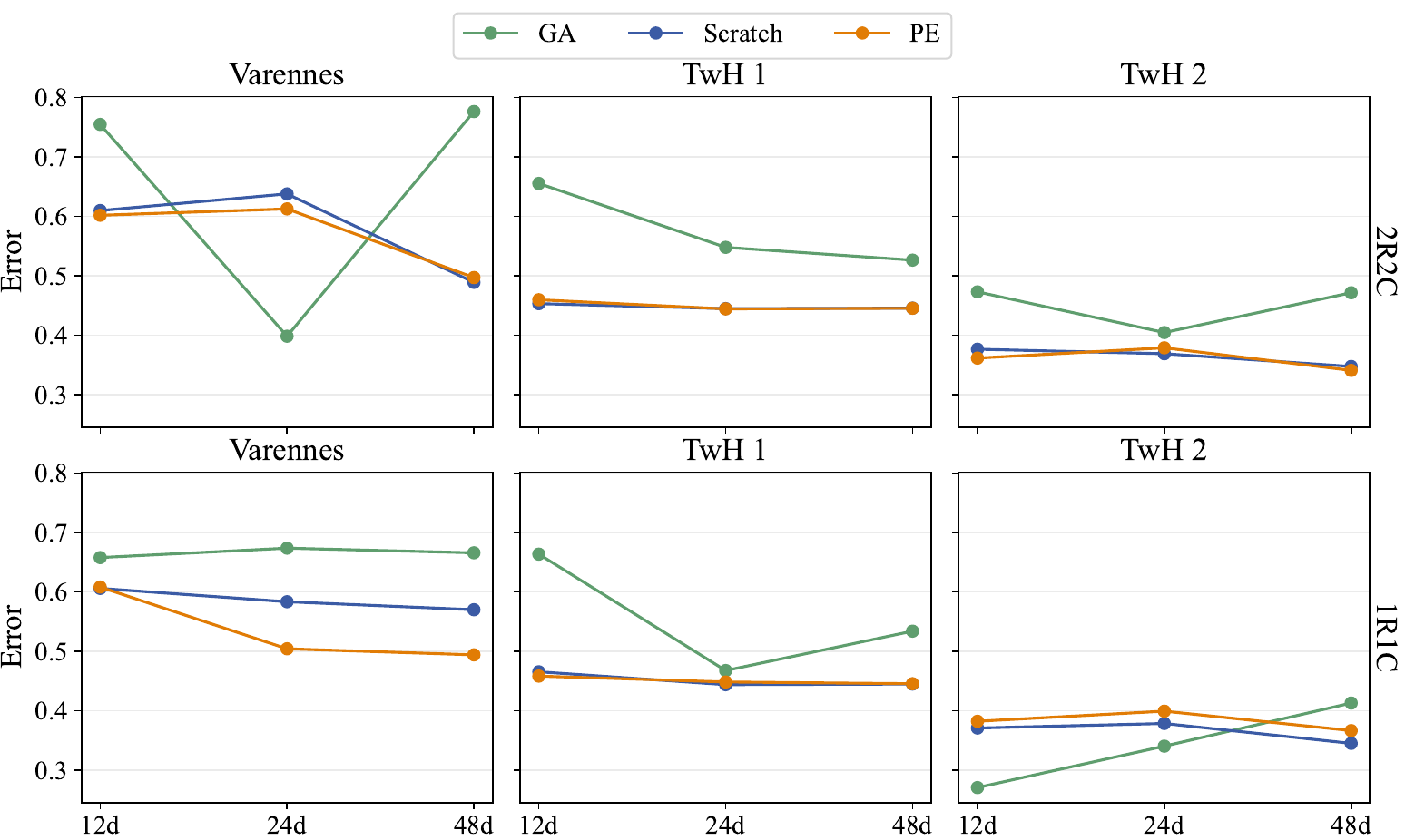}
    \vspace{-0.3cm}
    \caption{RMSE values in K as a function of training data length for the 2R2C and 1R1C models, shown for the real buildings.}
    \label{fig:indu_real}
\end{figure*}

\begin{figure*}[]
    \centering
    \includegraphics[width=0.85\linewidth]{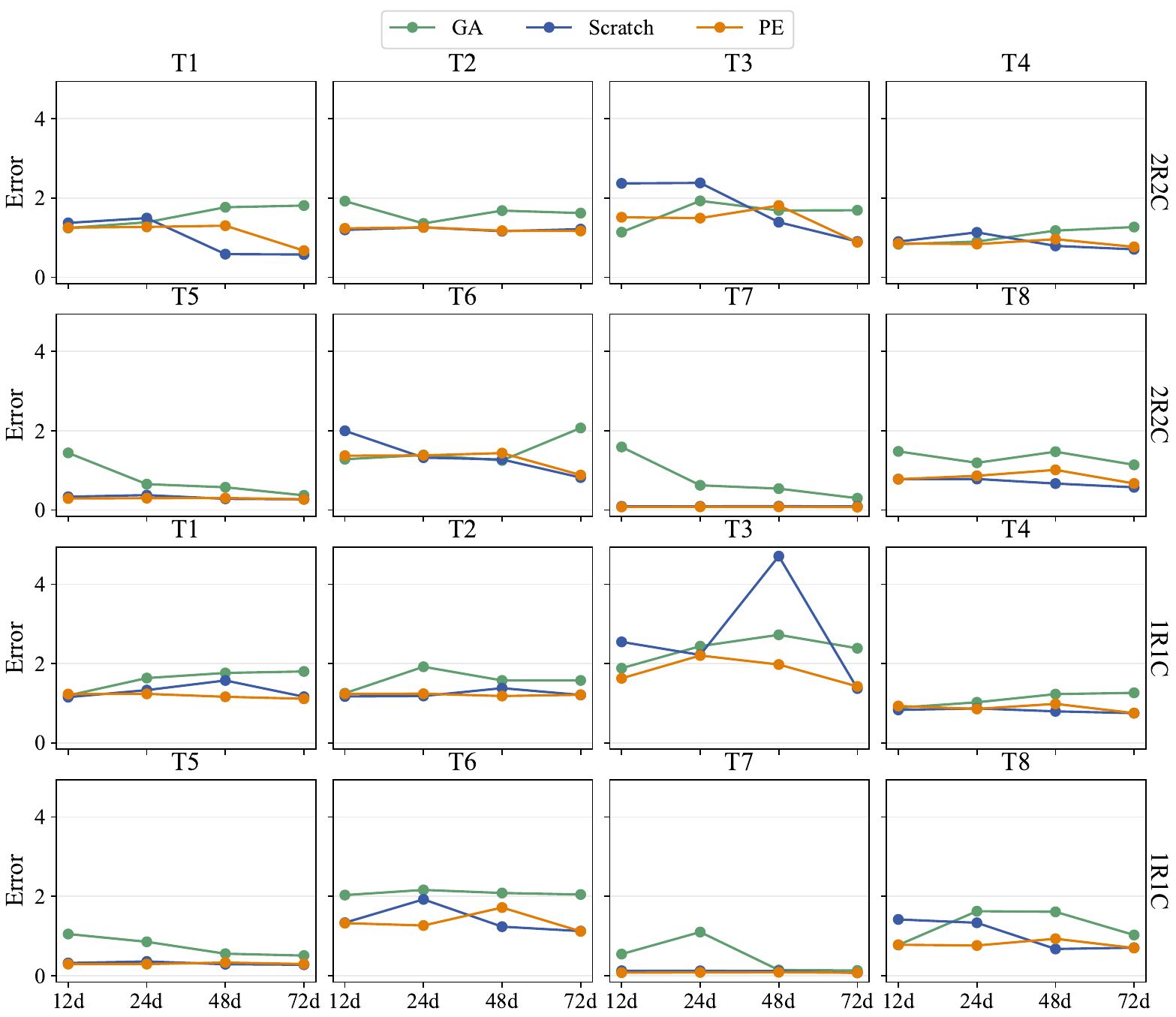}
    \vspace{-0.3cm}
    \caption{RMSE values in K as a function of training data length for the 2R2C and 1R1C models, shown for simulated buildings.}
    \label{fig:indu_builda}
\end{figure*}

Fig.~\ref{fig:indu_real} shows the individual results for the real-world buildings and Fig.~\ref{fig:indu_builda} for the simulated buildings. When considering up to 72 days of data for each method (i.e., the lowest RMSE reached at any point along the $y$-axis), the neural approaches achieve the strongest performance in 20 of 22 building-RC configuration combinations, for both simulated and real-world buildings. 

\section{Individual MPC results}
\label{indu_control}

\begin{figure}[H]
    \centering
    \includegraphics[width=1\linewidth]{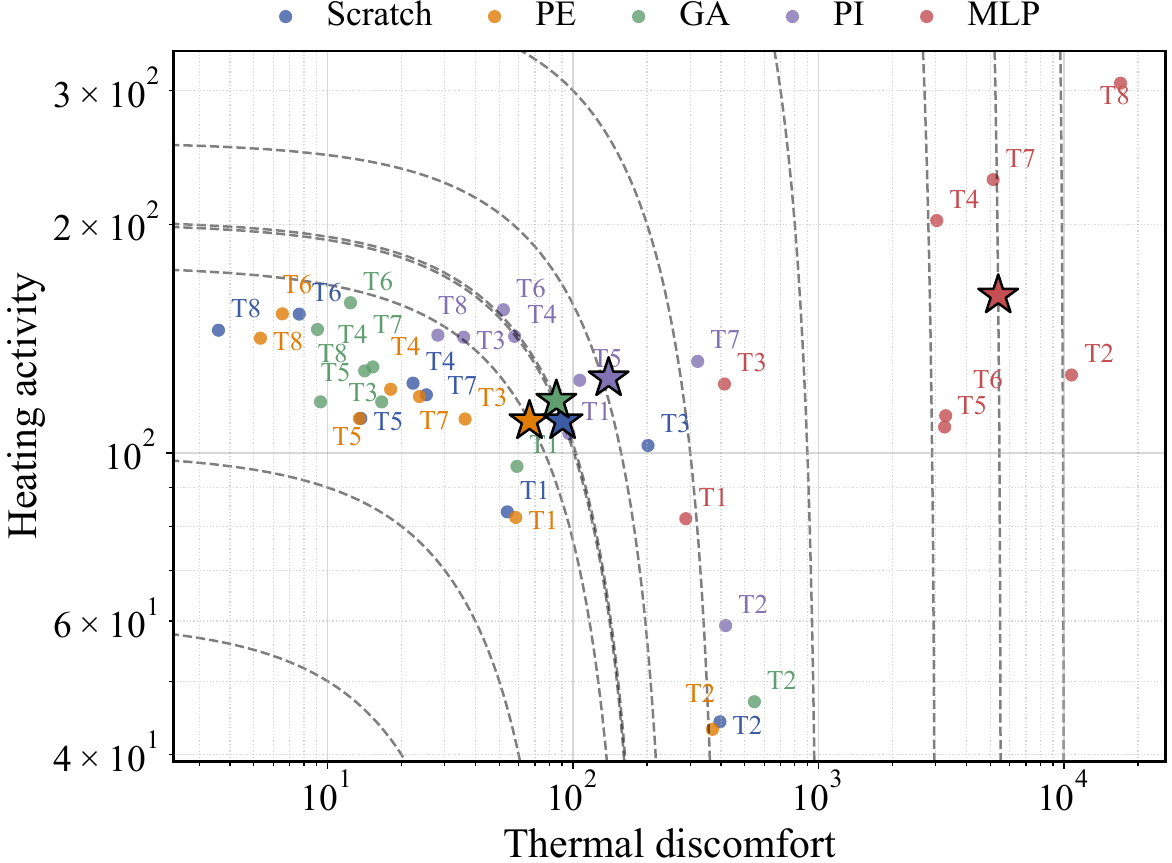}
    \vspace{-0.8cm}
    \caption{Heating activity vs.\ thermal discomfort per building and method (both axes log-scaled). Stars mark the mean per method; dashed lines indicate constant total cost.}
    \label{fig:mpc_scatter}
\end{figure}

Fig.~\ref{fig:mpc_scatter} breaks the total cost down into its two components, heating activity and comfort cost, for each building individually; the oscillation term ($w_2$) is omitted, as it is used only during optimization, not in the final cost. We use normalized heating (activity of $u$) rather than energy use, since nominal powers differ across buildings and would otherwise limit comparability. The three RC-based methods cluster tightly in the bottom-left region of low energy cost and low discomfort, with the Pretrained Estimator's mean (star) closest to the origin. The MLP, in contrast, is scattered across the top-right region for most buildings, indicating both high energy use and severe comfort violations. 

\end{document}